\documentclass[aps,prb,twocolumn,showpacs,letterpaper,noshowpacs]{revtex4}

\usepackage[T1]{fontenc}

\usepackage{amsmath,amssymb,amsfonts,upgreek,bm}
\usepackage{graphicx}

\graphicspath{{./}}

\begin{document}

\title{Spin transport and spin-caloric effects in (Cr,Zn)Te
  half-metallic nanostructures: Effect of spin disorder at elevated
  temperatures from first principles}

\date{\today}

\author{Roman Kov\'a\v{c}ik}
\email{r.kovacik@fz-juelich.de}
\affiliation{Peter Gr\"unberg Institut and Institute for Advanced
  Simulation, Forschungszentrum J\"ulich and JARA, 52425 J\"ulich, Germany}
\author{Phivos Mavropoulos}
\affiliation{Peter Gr\"unberg Institut and Institute for Advanced
  Simulation, Forschungszentrum J\"ulich and JARA, 52425 J\"ulich, Germany}
\author{Stefan Bl\"ugel}
\affiliation{Peter Gr\"unberg Institut and Institute for Advanced
  Simulation, Forschungszentrum J\"ulich and JARA, 52425 J\"ulich, Germany}

\begin{abstract}
    An important contribution to the thermoelectric and spin-caloric
    transport properties in magnetic materials at elevated
    temperatures is the formation of a spin-disordered state due to
    local moment fluctuations. This effect has not been largely
    investigated so far. We focus on various magnetic nanostructures
    of CrTe in the form of thin layers or nanowires embedded in ZnTe
    matrix, motivated by the miniaturization of spintronics devices
    and by recent suggestions that magnetic nanostructures can lead to
    extraordinary thermoelectric effects due to quantum
    confinement. The electronic structure of the studied systems is
    calculated within the multiple scattering screened
    Korringa-Kohn-Rostoker Green function (KKR-GF) framework. The
    Monte Carlo method is used to simulate the magnetization in the
    temperature induced spin disorder. The transport properties are
    evaluated from the transmission probability obtained using the
    Baranger-Stone approach within the KKR-GF framework. We find
    qualitative and quantitative changes in the thermoelectric and
    spin-caloric coefficients when spin-disorder is included in the
    calculation. Furthermore, we show that substitutional impurities
    in CrTe nanowires could considerably enhance the Seebeck
    coefficient and the thermoelectric figure of merit.
\end{abstract}


\maketitle

\section{Introduction}\label{sec:intro}


The rapidly growing field of spin
caloritronics\cite{2012-05//bauer/saitoh/van-wees} has in recent years
given impulse to diverse studies focusing on a coupling between
spintronics and thermoelectricity. The strong activity in the field is
reflected in numerous conceptual developments, including the
spin-Seebeck effect in
ferromagnets;\cite{2008-10//uchida/takahashi//saitoh} the
magneto-Seebeck effect in tunnel
junctions;\cite{2011-10//walter/walowski//heiliger} molecular
junctions\cite{2013-08//maslyuk/achilles//mertig} or
nanowires;\cite{2013-08//bohnert/vega//nielsch} the spin-orbit-based
anisotropy of the Seebeck coefficient;\cite{2013-09//popescu/kratzer}
the thermally induced spin accumulation at half-metal/normal metal
interfaces;\cite{2014-05//geisler/kratzer/popescu} the thermal
spin-transfer torque;\cite{2013-03//leutenantsmeyer/walter//heiliger}
the spin-based Peltier
cooling;\cite{2005-01//fukushima/yagami//yuasa,2010-06//sugihara/kodzuka//fukushima,2011-01//vu/sato/katayama-yoshida,2012-02//scharf/matos-abiague//fabian}
the effect of spin disorder on the transport coefficients
\cite{2014-04//kovacik/mavropoulos//blugel} and the
transverse-transport effects (spin Nernst and anomalous Nernst
effect);\cite{2012-07//tauber/gradhand//mertig,2013-11//wimmer/kodderitzsch//ebert,2013-02//weischenberg/freimuth//mokrousov}
among other studies.

A strong potential for the design of materials with tailored
spin-caloric properties lies in the realm of magnetic nanostructures,
which are at the focus of the present paper. The advantages of
nanostructured materials and junctions are well known. In the first
place, quantum confinement effects are quite pronounced, giving the
possibility of electronic structure design by means of geometrical
design. Additionally, targeted novel-phase design is possible by
out-of-equilibrium growth that can be easily sustained in
nanostructures compared to the bulk.  Another advantage is that spin
transfer is less volatile since the electron spin relaxation length
can be larger than the nanostructure diameter.  These ideas have been
discussed in several works, as follows.

Firstly, the possibility of an enhanced Peltier effect in
submicron-sized metallic junctions was
suggested.\cite{2005-01//fukushima/yagami//yuasa,2010-06//sugihara/kodzuka//fukushima,2011-01//vu/sato/katayama-yoshida}
Secondly, a thermoelectric-cooling mechanism was theoretically
proposed, based on the adiabatic spin-entropy expansion in a
quasi-one-dimensional nano-superstructure (the so called ``Konbu''
phase) by injecting a spin current from a ferromagnetic to a
paramagnetic metal.\cite{2007-09//katayama-yoshida/fukushima//sato}
The realization of Konbu phases as a result of spinodal
nano-decomposition was extensively studied in wide gap diluted
magnetic
semiconductors.\cite{2005-07//sato/katayama-yoshida/dederichs,2006-04//fukushima/sato//dederichs,2007-03//katayama-yoshida/sato//dederichs,2007-07//sato/fukushima/katayama-yoshida}
Effective chemical pair interactions between magnetic impurities were
found to be attractive and short ranged, suggesting a tendency towards
the phase separation.\cite{2006-04//fukushima/sato//dederichs} An
increase of crossover temperature $T_\text{c}$ (either blocking or
Curie temperature) results due to the formation of magnetic
percolation paths.\cite{2005-07//sato/katayama-yoshida/dederichs} A
new crystal growth method of positioning by seeding and shape
controlling was proposed, with $100~\text{Tb/in}^2$ density of
high-$T_\text{c}$ nano-magnets in the semiconductor
matrix.\cite{2007-03//katayama-yoshida/sato//dederichs}

Regarding the spin-polarized transport, one point that has received
little attention is the effect of magnetic moment fluctuations at
elevated temperatures, known as \emph{spin disorder}. The well-known
effect of spin-disorder-induced
resistivity\cite{1964-09//arajs/colvin,1967-07//kierspe/kohlhaas/gonska,2009-12//wysocki/sabirianov//belashchenko,2008-02//buruzs/szunyogh/weinberger,2012-06//glasbrenner/belashchenko//turek}
has also its counterparts in the current spin polarization and in the
Seebeck and spin-Seebeck effects. While in the bulk of magnetic
materials the moment fluctuations are small for temperatures
significantly lower than $T_\text{c}$ (even up to
$\frac{2}{3}T_\text{c}$), in nanostructures the surface-to-volume
ratio is large and the fluctuations are accordingly stronger. In a
previous work,\cite{2014-04//kovacik/mavropoulos//blugel} we reported
calculations on spin transport influenced by spin-disorder effects in
cobalt nanostructures embedded or not in a copper matrix. In the
present study, we shift our focus towards half-metallic materials,
where the absence of states in the spin-down channel can act in favor
of very pronounced spin-transport effects. Such effects have been
studied in the field of spin caloritronics in the case of
half-metallic Heusler
alloys;\cite{2010-02//barth/fecher//kobayashi,2013-09//popescu/kratzer,2014-05//geisler/kratzer/popescu,2014-03//comtesse/geisler//szunyogh}
however, no investigation of the spin disorder effect at elevated
temperatures has so far been conducted.

In the present report we address this issue, focusing on the effect of
temperature-induced spin disorder on the charge- and spin-current,
thermopower, and spin-Seebeck coefficient. Instead of Heusler alloys,
we choose CrTe as a model system, as previous studies suggest it can
be grown in nanostructured form, within a ZnTe matrix, retaining its
half-metallic character (see below).  We calculate the transport
properties of CrTe thin films and CrTe wires of single-atom cross
section (``monoatomic'') embedded in ZnTe. The motivation for choosing
monoatomic wires as test systems is that quantum confinement effects
should be most pronounced in such structures, which could be
fabricated by seed-induced molecular beam
epitaxy\cite{2007-03//katayama-yoshida/sato//dederichs} as a limiting
case of the Konbu-phase.  We work within the Landauer-B\"uttiker
approach, employing the multiple scattering screened
Korringa-Kohn-Rostoker Green function (KKR-GF)
framework,\cite{2004-03//mavropoulos/papanikolaou/dederichs,2006-11//yavorsky/mertig,2014-04//kovacik/mavropoulos//blugel}
with a realistic treatment of the real-space spin disorder at finite
temperatures using a
supercell.\cite{2014-04//kovacik/mavropoulos//blugel}

This paper is organized as follows. Section~\ref{sec:previous}
contains a summary of previous works on CrTe. In Sec.~\ref{sec:met} we
define the studied model systems, describe the methods used in this
work and corresponding computational details. Results are discussed in
Sec.~\ref{sec:res} and the conclusions are given in
Sec.~\ref{sec:sum}.

\section{Summary of investigations on C\lowercase{r}T\lowercase{e} and (C\lowercase{r},Z\lowercase{n})T\lowercase{e} nanostructures\label{sec:previous}}


A promising material to exhibit the Konbu phase formation is the
(Cr,Zn)Te diluted magnetic semiconductor (DMS), composed of a ZnTe
matrix with a Cr impurity concentration of about
5\%.\cite{2006-04//fukushima/sato//dederichs} ZnTe is a semiconductor
with a direct band gap of around
${2.3~\text{eV}}$,\cite{1957-11//larach/shrader/stocker,1963-07//cardona/greenaway,1984-12//venghaus}
normally crystallizing in the zinc blende (ZB) crystal structure. CrTe
in its various phases has been widely studied in the past.  It is a
ferromagnetic transition metal chalcogenide with
${T_\text{c}=334~\text{K}}$.\cite{1960-11//hirone/chiba} Although its
most stable crystal structure is that of the hexagonal NiAs type, it
can be grown on various different surfaces adapting to the underlying
cubic crystal
structure.\cite{1972-08//goswami/nikam,2007-09//sreenivasan/teo//osipowicz}
Theoretical investigations have predicted half-metallicity in the
ZB-CrTe structure.\cite{2003-03//galanakis/mavropoulos}


Electrical and magnetic properties of CrTe were experimentally studied
in the bulk form or often in its stable stoichiometric
Cr$_{N-1}$Te$_N$ compounds with ordered Cr vacancies. The
ferromagnetic phase was found stable under hydrostatic pressure up to
$2.5~\text{GPa}$, where the $T_\text{c}$ dropped to ${170~\text{K}}$
linearly from its ambient pressure
value.\cite{1978-02//lambert-andron/grazhdankina/vettier} A saturation
of the spin disorder contribution to the electrical resistivity $\rho$
in the NiAs structure was observed in the vicinity of $T_\text{c}$ as
a characteristic kink in
$\rho(T)$.\cite{1961-08//grazhdankina/gaidukov//shchipanov,1966-02//nogami,1989-11//dijkstra/weitering//de-groot}
Seebeck coefficient measurements were reported for stoichiometric
Cr$_3$Te$_4$\cite{1981-02//peix/babot/chevreton} and
non-stoichiometric Cr$_{0.8}$Te and
Cr$_{0.9}$Te\cite{1989-11//dijkstra/weitering//de-groot} compounds in
the NiAs crystal structure, showing a non-trivial dependence on the
Cr/Te ratio.

Room temperature (RT) ferromagnetism (${T_\text{c} \approx
300~\text{K}}$) was reported in Zn$_{1-x}$Cr$_{x}$Te semiconducting
thin films with ${x=0.2}$.\cite{2003-05//saito/zayets//ando} In a more
recent study, structural, magnetic and transport properties of the
(Cr,Zn)Te compound system were
investigated.\cite{2007-09//sreenivasan/teo//osipowicz} The
selected-area electron diffraction method indicated presence of
ZB-CrTe coherent with the ZB-ZnTe buffer in both the CrTe thin films
and in the CrTe nanocluster precipitates, having a $T_\text{c}$ of
${247~\text{K}}$ and ${220~\text{K}}$, respectively. Superparamagnetic
behavior was found in the CrTe thin films (Zn$_{1-x}$Cr$_{x}$Te with
${x>0.12}$) which further showed metallic conductance (${\rho \approx
10~\upmu\Omega~\text{m}}$ at RT), in agreement with previously
published results.\cite{1961-08//grazhdankina/gaidukov//shchipanov}
The CrTe precipitates (${x=0.14}$) exhibited dirty metallic-like
character (${\rho \approx 10^{3}~\upmu\Omega~\text{m}}$ at RT) with
the $\rho(T)$ spin-disorder kink coinciding with
$T_\text{c}$. Zn$_{1-x}$Cr$_x$Te samples with ${x \le 0.12}$ were
found to be highly resistive with ${\rho \approx
10^{7}~\upmu\Omega~\text{m}}$. Further on, the electronic structure of
Zn$_{1-x}$Cr$_{x}$Te DMS was investigated by x-ray magnetic circular
dichroism and photoemission spectroscopy. It was concluded that
ferromagnetism originates from Cr ions of a single chemical
environment with a spatially isotropic electronic
configuration.\cite{2008-05//kobayashi/ishida//ando}


Numerous ab-initio studies on the bulk properties of CrTe have been
reported.\cite{2003-03//galanakis/mavropoulos,2003-07//xie/xu//pettifor,2007-08//mavropoulos/galanakis,2010-09//liu/bose/kudrnovsky,2010-10//vadkhiya/dashora//ahuja,2011-02//guo/liu}
The stability of the half-metallicity was studied in the case of bulk
ZB-CrTe in the experimental lattice parameter of several
semiconductors,\cite{2003-03//galanakis/mavropoulos} as well as in
both Cr and Te terminated (001) surfaces of
ZB-CrTe,\cite{2008-07//yun/hong} non-stoichiometric cubic binary
chromium chalcogenides\cite{2009-12//guo/liu} and Cr/Mn chalcogenide
interfaces.\cite{2008-04//nakamura/akiyama//freeman} A magnetic phase
diagram of CrTe based on the Korringa-Kohn-Rostoker ab-initio
calculations and Monte Carlo simulations was reported showing good
agreement between theory and
experiment.\cite{2010-04//polesya/mankovsky//bensch} Recently, the
CrTe(001)/ZnTe(001) interface was studied with ab-initio methods
showing a coherent change from ferromagnetic half-metal to nonmagnetic
insulator.\cite{2010-04//ahmadian/abolhassani//elahi} Further on, the
stability of ferromagnetism in quasi-one-dimensional Cr chains
embedded in ZnTe was reported.\cite{2013-07//nakayama/fujita/raebiger}

\section{Method and computational details}\label{sec:met}

\subsection{Geometric setup of the model systems\label{ssec:met-setup}}

\begin{figure}
  \includegraphics[width=\columnwidth]{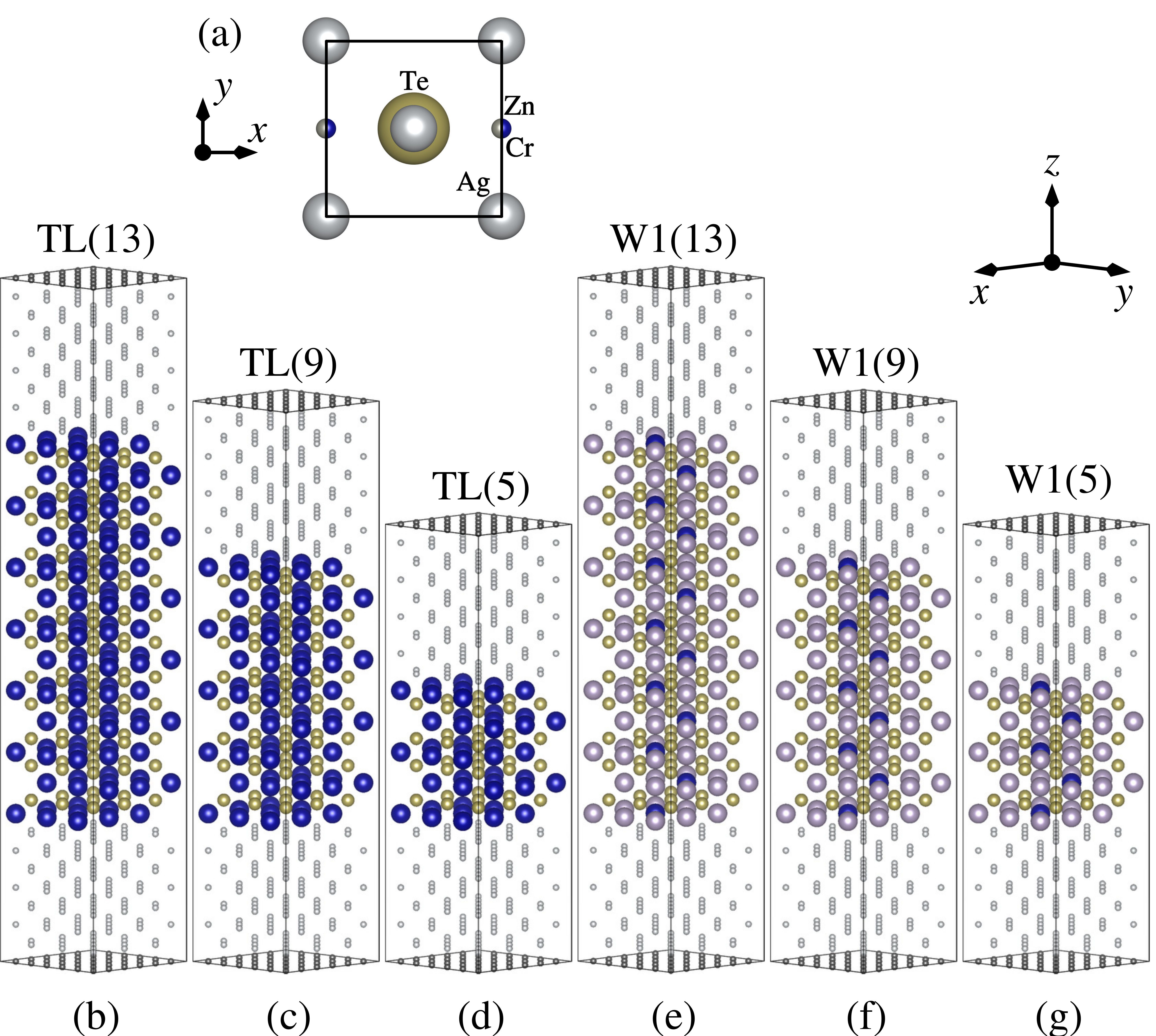}
  \caption{(Color online) (a) Ag/(Cr,Zn)Te interface unit cell, the
    layers are ordered along the $z$ direction as Te, Cr/Zn and
    Ag. CrTe thin layers (TL) (b)-(d) and monoatomic Cr nanowires (W1)
    embedded in ZnTe matrix (e)-(g) sandwiched between Ag leads
    (oriented according to the $xyz$ compass). Different elements in
    (b)-(g) are depicted as large dark spheres (Cr), large bright
    spheres (Zn), medium size spheres (Te) and small spheres (Ag). The
    number of Cr layers is 13 [(b),(e)], 9 [(c),(f)], and 5
    [(d),(g)]. The structure representations were plotted using
    VESTA.\cite{2011-12//momma/izumi}}
  \label{fig:str}
\end{figure}

Our model systems are composed of a central region in the zinc blende
crystal structure (experimental lattice parameter of ZB-ZnTe
${a_\text{lat}=6.1~\text{\AA}}$) sandwiched between half-infinite Ag
leads extending in the $\pm z$-directions (Fig.~\ref{fig:str}). The
central region is formed either by a thin layer (TL) of CrTe or by a
monoatomic CrTe wire (W1) embedded in the ZnTe matrix, with Zn/Cr
atoms interfacing the leads. The in-plane unit cell of the interface
with lattice constant $a_\text{lat}/\sqrt{2}$ is shown in
Fig.~\ref{fig:str}(a). The Ag atoms at the interface assume the
positions of the Te and vacant sites in the ZB structure. Due to the
5\% mismatch between the face-centered cubic Ag lattice parameter
({4.1~\AA}) and $a_\text{lat}/\sqrt{2}$, the natural lattice of Ag
atoms is compressed in the $z$ direction in order to preserve the Ag
unit cell volume. A ${3 \times 3}$ in-plane supercell is used to model
the real-space spin disorder in the CrTe thin layers
[Figs.~\ref{fig:str}(b)--\ref{fig:str}(d)] and as separation between
the in-plane periodic images of the nanowires
[Figs.~\ref{fig:str}(e)--\ref{fig:str}(g)]. The central region
thickness of 13, 9 or 5 Cr layers is chosen to examine the influence
of a quasi-3D to 2D transition (in TL systems) or a quasi-1D to 0D
transition (in W1 systems) on the transport properties.

\subsection{Terminology \label{ssec:termin}}

A few words on the terminology of later sections are due. We use the
term \emph{core region} for the CrTe/ZnTe region between the Ag leads
in the structures, i.e., either the CrTe film or the nanowire with the
embedding matrix. By \emph{interface} we mean the interface between
the Ag leads and the core region. We further indicate by a number in
curly brackets the distance of a given atom to the interface in $z$
direction, e.g, Cr\{7\} is the seventh Cr atom counting from the
interface and Ag\{1\} is the interface Ag atom. Round brackets are
used to declare the thickness of the core region, e.g., W1(13) is the
system with the 13-atom-long CrTe nanowire.

\subsection{Self-consistent calculations}\label{ssec:met-calc}

The electronic structure of the model systems was calculated by the
KKR-GF method using the full-potential
formalism\cite{2002-03//papanikolaou/zeller/dederichs,1990-09//stefanou/akai/zeller,1991-09//stefanou/zeller,SPR-TB-KKR}
and the local density approximation\cite{1980-08//vosko/wilk/nusair}
to the exchange-correlation energy functional. The angular momentum
expansion was truncated at ${l_\text{max}=3}$. All structural
parameters were kept fixed and the supercell potentials were
constructed following the procedure described in Section II.B of
Ref.~[\onlinecite{2014-04//kovacik/mavropoulos//blugel}]. A
well-converged density was reached by using an ${18 \times
18~k}$-point mesh for the integration in the unit cell surface
Brillouin zone (SBZ) and a smearing electronic temperature of
${800~\text{K}}$.\cite{1995-10//wildberger/lang//dederichs}

It is well known that the local density approximation to
density-functional theory underestimates the band gap of semiconductors and
insulators. This can pose a practical problem in the
transport calculations in systems with ZnTe, increasing the tunneling
amplitude to unrealistically high values that could play a spurious
role in the thin-spacer systems. We circumvent the problem by
adjusting the self-consistent atomic potentials of Zn and Te (by rigid
positive shift of the Zn potential and rigid negative shift of the Te
potential) such that the band gap value corresponds to the
experimental one\cite{1984-12//venghaus} and the Fermi energy is set
in its middle.

\subsection{Spin disorder and electron transport\label{ssec:met-sdet}}

The calculation method used in this study was in detail described in
Ref.~\onlinecite{2014-04//kovacik/mavropoulos//blugel}. However, for
completeness, the key steps of the procedure are outlined in this
section, where also the details specific to the studied systems are
given.

The spin disorder model is based on adopting the moment directions at
temperature $T$ as they are given by a classical Heisenberg model
\begin{equation}\label{eq:hh}
  H=-\sum_{i,j}J_{ij}\ \mathbf{M}_i\cdot\mathbf{M}_j.
\end{equation}
Here, $\mathbf{M}_i$ and $\mathbf{M}_j$ are unit vectors pointing in
the direction of the magnetic moments at sites $i$ and $j$,
respectively, while $J_{ij}$ are the exchange parameters extracted
from the ground-state electronic structure by means of the method of
infinitesimal rotations.
\cite{1987-05//liechtenstein/katsnelson//gubanov} The moment
directions are given in a series of sampling configurations
(``snapshots'') at thermal equilibrium at $T$ by means of the Monte
Carlo (MC) method employing the Metropolis algorithm
\cite{1953-06//metropolis/rosenbluth//teller} and using the Mersenne
twister \cite{1998-01//matsumoto/nishimura} for the random number
generation. The number of MC sampling configurations $N_{\text{conf}}$
necessary for a statistical convergence of the transport properties
was chosen proportional to the fluctuation amplitude as described in
Ref. \onlinecite{2014-04//kovacik/mavropoulos//blugel}, yielding
typical ${N_\text{conf} \approx 300}$ in the vicinity of $T_\text{c}$,
where the moment fluctuations are largest.

The spin-up (majority-spin) and spin-down (minority-spin) directions
in the electronic structure and transport calculations for each MC
snapshot are calculated with respect to the global magnetization axis
of the same MC snapshot and averaged at the end over all snapshots.

We further define the moment-moment correlation function $C_N$ between
moments, averaged over all layer pairs having a distance of $N$ layers
(distance of ${Na_\text{lat}/2}$ in the $z$ direction). For a given
$N$, let there be $N^{\rm pair}$ of such pairs in the nanostructure;
also, let the first of the layers in the pair be indicated by
${c=1,2,\ldots}$, with the in-plane coordinates of the atoms indicated
by $(a,b)$; this places the second layer at ${c+N}$ with in-plane
coordinates $(a',b')$. The in-plane supercell contains $N_{ab}$ (${N_a
\times N_b}$) magnetic atoms.  Then we define
\begin{equation}\label{eq:zcorr}
  C_{N}(T)=
  \frac{1}{N_{ab}}
  \sum_{a,b}
  \frac{1}{N^{\rm pair}}
  \sum_{c}\langle{\mathbf{M}_{(a,b);c}\cdot\mathbf{M}_{(a',b');c+N}}\rangle_T
  .
\end{equation}
In the calculation of $C_N$ we include correlations between pairs
$(a,b)$ and $(a',b')$ whose distance, when projected onto the $xy$
plane, does not exceed ${a_\text{lat}/\sqrt{2}}$.

In calculating the transmission, we work within the adiabatic
approximation,\cite{1996-07//antropov/katsnelson//kusnezov,1998-07//halilov/eschrig//oppeneer}
assuming that the electrons do not exchange energy with the magnetic
system while traversing the nano-sized junction and that the magnetic
moments can be treated as frozen in their magnitude and direction
during this short time interval.  The same conceptual real-space
approach was applied for the spin-disorder resistivity of
ferromagnets,\cite{2009-12//wysocki/sabirianov//belashchenko,2012-06//glasbrenner/belashchenko//turek}
but with the difference that in those works only the high temperature
limit of the paramagnetic state (complete spin disorder) was
considered. The transmission probability matrix through the
non-collinear magnetic structure is calculated by the same code that
was developed for
Ref.~\onlinecite{2014-04//kovacik/mavropoulos//blugel}; it is based on
a combination of the Baranger-Stone\cite{1989-10//baranger/stone}
Green function approach to the Landauer-B\"uttiker theory and on the
KKR-GF
method.\cite{2002-03//papanikolaou/zeller/dederichs,2004-03//mavropoulos/papanikolaou/dederichs}

A further approximation is that we accept a rotation of the
ground-state magnetic part of the site-dependent potentials in the
instantaneous direction prescribed by the MC. This is done without a
self-consistent calculation of the non-collinear state, which would
result in an increase of computational time by one to two orders of
magnitude. We verified that the magnitude of the magnetic moments in
the non-collinear state differ very little from their respective
values at the ground state. From this we conclude that the
intra-atomic exchange interaction is dominant over the inter-atomic
interaction for the moment formation, justifying the
non-self-consistent approximation.

The basic quantities in (spin)-thermoelectric calculations are the
well-known transport coefficients $\mathbf{L}_n$ (here, boldface font
implies a $2\times 2$ matrix in spin space). These were evaluated by a
numerical integration of the transmission probability
$\boldsymbol{\Gamma}(\mathbf{k}_{\parallel},E)$\cite{2014-04//kovacik/mavropoulos//blugel}
[which is also a matrix in spin space with elements $\varGamma^{\sigma
\sigma'}(\mathbf{k}_{\parallel},E)$] over the crystal momentum
$\mathbf{k}_{\parallel}$ and energy $E$ as
\begin{equation}\label{eq:Ln}
  \mathbf{L}_n=
  -\int \text{d}E \,
  (E-E_\text{F})^n \,
  \frac{\partial f_T(E)}{\partial E}
  \int_{\text{SBZ}} \! \text{d}\mathbf{k}_{\parallel}
  \boldsymbol{\Gamma}(\mathbf{k}_{\parallel},E).
\end{equation}
Here,
${f_T(E)=\left[\text{exp}\left(\frac{E-E_\text{F}}{k_\text{B}T}\right)+1\right]^{-1}}$
is the Fermi-Dirac distribution function, $T$ is the temperature of
the MC simulation, and $E_\text{F}$ is the Fermi energy. For each
temperature and system,
${\boldsymbol{\Gamma}(E)=\int_{\text{SBZ}}\text{d}\mathbf{k}_{\parallel}\boldsymbol{\Gamma}(\mathbf{k}_{\parallel},E)}$
was calculated on a mesh of 15 equidistant points in the range
${-7\,k_\text{B}T\leq(E-E_\text{F})\leq +7\,k_\text{B}T}$, beyond
which ${(E-E_\text{F})^n\partial{f_T}/\partial{E}}$ practically
vanishes. Tests on denser grids gave insignificant
differences. Finally, $\langle \mathbf{L}_n \rangle_T$ is calculated
as an average over the non-collinear MC configurations.

The electrical conductance $G$, electrical resistance $R$, charge
Seebeck coefficient $S_{\text{C}}$ and spin Seebeck coefficient
$S_{\text{S}}$, the thermal conductance $K$, and the thermoelectric
figure of merit $ZT$ are calculated using the well-known formulas
\begin{align}
  G^{\sigma\sigma'}(T)&=\frac{e^2}{h}\langle {L}_0^{\sigma\sigma'}\rangle_T \label{eq:Gss}\\
  G&={\textstyle\sum_{\sigma\sigma'}G^{\sigma\sigma'}} \label{eq:G}\\
  R&=\frac{1}{G} \label{eq:R}\\
  S_\text{C}&=-\frac{\sum_{\sigma\sigma'}\langle L_1^{\sigma\sigma'}\rangle_T}
  {eT\sum_{\sigma\sigma'}\langle L_0^{\sigma\sigma'} \rangle_T} \label{eq:SC}\\
  S_\text{S}&=-\frac{
    \langle L_1^{\uparrow\uparrow}\rangle_T+\langle L_1^{\downarrow\uparrow}\rangle_T
    -\langle L_1^{\downarrow\downarrow}\rangle_T-\langle L_1^{\uparrow\downarrow}\rangle_T
  }
  {eT\sum_{\sigma\sigma'}\langle L_0^{\sigma\sigma'}\rangle_T} \label{eq:SS}\\
  K&=\frac{
    \sum_{\sigma\sigma'}\langle L_2^{\sigma\sigma'}\rangle_T
    -\left(\sum_{\sigma\sigma'}\langle L_1^{\sigma\sigma'}\rangle_T\right)^2
  }
  {T\sum_{\sigma\sigma'}\langle L_0^{\sigma\sigma'} \rangle_T} \label{eq:K}\\
  ZT&=\frac{GT}{K}S_\text{C}^2 \label{eq:ZT}
  .
\end{align}

\section{Results and discussion}\label{sec:res}

\subsection{Electronic structure of the collinear magnetic state\label{ssec:res-est0}}

Before we proceed to the analysis of the transport properties, we
briefly discuss the electronic structure of the model systems before
imposing temperature on the magnetic system and non-collinear
magnetism.  The self-consistent calculations were performed assuming a
collinear magnetic ground state. However, a detailed inspection of the
exchange coupling parameters (not shown here) already indicates that
in the thin layers the interface Cr moments tend to a non-collinear
ground state. Monte Carlo simulations indeed show that while the
film-interior magnetization, i.e., deeper than in the interface layer,
remains practically collinear, moments at the interface tilt away from
the film-interior magnetization but not fully reaching the spin-flop
state. Their projection to the plane normal to the film-interior
magnetization forms a checkerboard pattern. The non-collinearity at
low temperatures is accounted for in the conductance calculations (see
next Sec.~\ref{ssec:res-sd} for more details). Wires, on the other
hand, show a collinear ground state.

\begin{figure}
  \includegraphics[width=\columnwidth]{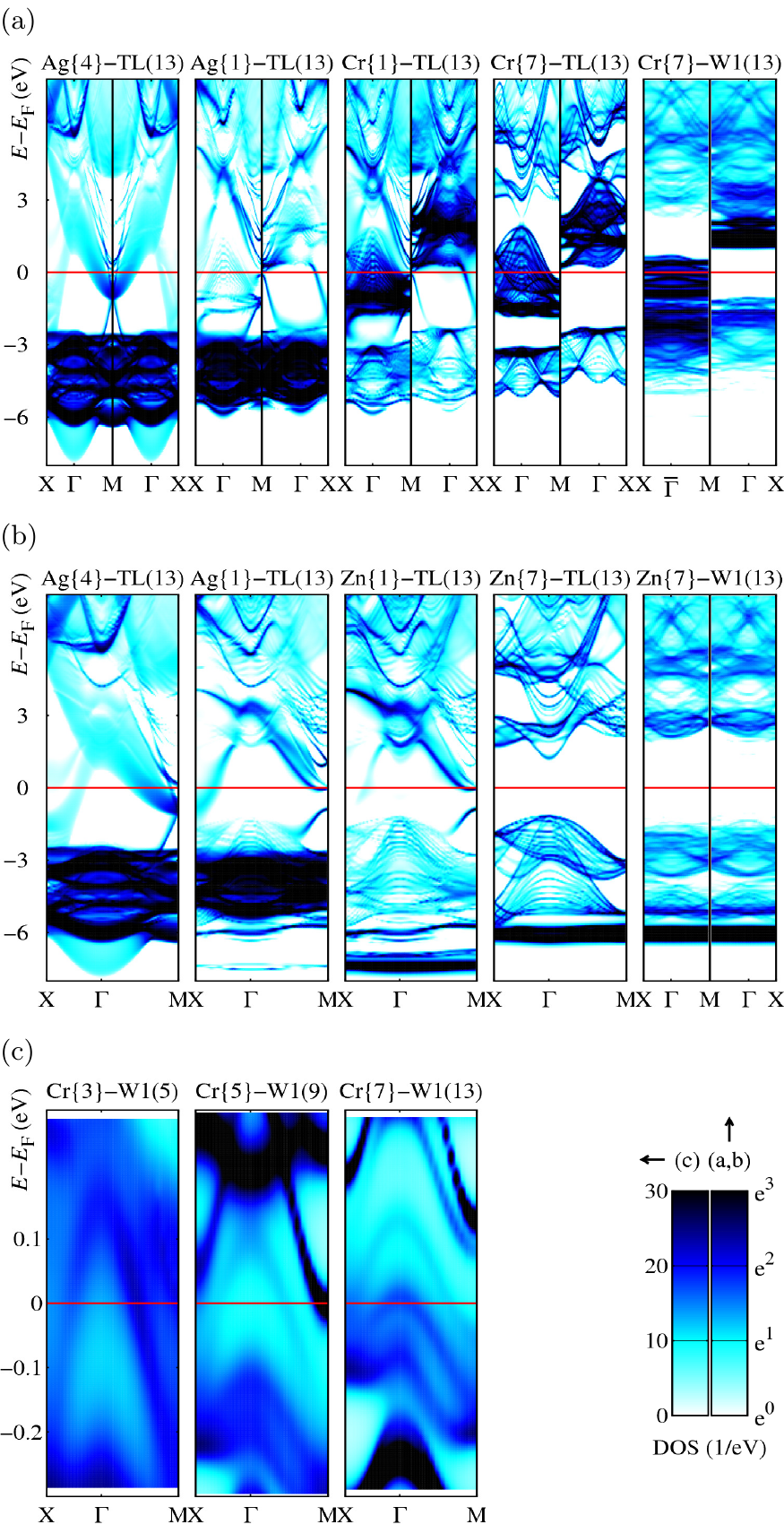}
  \caption{(Color online) The $\mathbf{k}$-resolved density of states
  (DOS) along the $\text{X-}\Gamma\text{-M}$ path in the surface
  Brillouin zone. The first four panels in (a) and (b) correspond to
  the total DOS of selected sites in the Ag/CrTe/Ag (a) and Ag/ZnTe/Ag
  (b) systems with ${1 \times 1}$ unit cell cross-section and 13 Cr
  (or Zn) layers. The number in curly brackets in the graph label
  denotes the distance of the labeled site from the interface. The
  rightmost panel in (a) and (b) corresponds to the DOS of the Cr site
  in the nanowire and the Zn site farthest from the Cr nanowire,
  respectively, in the central layer of the W1(13) system. Panels
  divided by a vertical line show the majority/minority DOS in their
  left/right part. (c) $\mathbf{k}$-resolved DOS at the central Cr
  atom of the W1(5), W1(9) and W1(13) nanowires in a narrow energy
  range around the Fermi level (indicated by horizontal line at
  ${E=0}$ in all graphs of this figure).}
  \label{fig:qdos}
\end{figure}

The $\mathbf{k}_\parallel$-resolved density of states (DOS) along the
$\text{X-}\Gamma\text{-M}$ path in the surface Brillouin zone is shown
for selected sites and systems in Fig.~\ref{fig:qdos}. Each panel in
Fig.~\ref{fig:qdos}(a) is divided in left/right sub-panels for
majority/minority spin channels. The first four panels from left in
Fig.~\ref{fig:qdos}(a) correspond to the TL(13) system with a ${1
\times 1}$ unit cell cross-section, i.e., the system has one atom per
layer in the two-dimensional unit cell in the CrTe part of the slab,
as shown in Fig.~\ref{fig:str}(a). The DOS at the Cr\{7\} atom,
located at the center of the core region, is essentially bulk-like,
characterized by the $p$ bands of Te at 2 to ${5~\text{eV}}$ below
$E_\text{F}$, partially occupied majority spin $d$ bands of Cr and the
bottom edge of the minority spin Cr $d$ bands just above
$E_\text{F}$. Similarly, the DOS of Ag\{4\} (fourth Ag atomic layer
from the interface) already shows bulk-like behavior, with the lower
edge of the characteristic $s$ band at $8~\text{eV}$ below
$E_\text{F}$ crossed by $d$ character bands between $6$ and
$2.5~\text{eV}$ below $E_\text{F}$ and with the DOS at $E_\text{F}$
dominated by an $sp$ band. The DOS at the interface Ag\{1\} and
Cr\{1\} atoms shows well preserved $d$ bands of the corresponding
element. Interface states of $sp$ character appearing just above
$E_\text{F}$ are completely suppressed in the Ag\{7\} layer (not
shown).

The first four panels from left in Fig.~\ref{fig:qdos}(b) correspond
to the system which is geometrically identical to the previously
discussed one but the core region consists of ZnTe instead of CrTe and
serves as a matrix for the W1(13) nanowire. The DOS at the Zn\{7\}
atom, located in the center of the slab, is characterized by fully
occupied $d$ bands of Zn at $6~\text{eV}$ below $E_\text{F}$ and the
Te $p$ bands right above them, separated from the Zn $s$ bands by a
direct band gap at the $\Gamma$ point. Note that the ZnTe gap was
manipulated to reach the experimental value as described in
Sec.~\ref{ssec:met-calc}. The DOS of Ag\{4\} is very similar to that
of Fig.~\ref{fig:qdos}(a) and likewise, the $sp$ interface state is
quickly suppressed with essentially no trace present in the Ag\{7\}
layer (not shown).

The rightmost panels in Fig.~\ref{fig:qdos}(a) and
Fig.~\ref{fig:qdos}(b) show the DOS of the Cr site in the nanowire and
the Zn site farthest from the Cr nanowire (second nearest in-plane
neighbor), respectively, in the central layer of the W1(13) system. As
can be seen in the Cr\{7\}-W1(13) panel of Fig.~\ref{fig:qdos}(a), the
dispersion of Cr $d$ bands is narrowed down while it is widened for
the Te $p$ bands due to the strong influence of the ZnTe matrix
[compare to the Zn\{7\}-TL(13) panel in Fig.~\ref{fig:qdos}(b)]. As a
result, the minority spin band gap is slightly smaller and shifted
upwards. The Zn\{7\}-W1(13) panel in Fig.~\ref{fig:qdos}(b) shows that
the semiconducting nonmagnetic character of ZnTe DOS is well preserved
in the ZnTe matrix surrounding the Cr nanowire.

While the difference between the DOS of the individual TL systems in
their central layer is rather small, the same is not true in the case
of the nanowires. Figure~\ref{fig:qdos}(c) displays the Cr atom DOS of
the nanowire central layer in a narrow energy range around
$E_\text{F}$, where the dominant contributions to the transport
properties are calculated. A characteristic pattern can be recognized
for all three systems, with strongly smeared-out features in case of
the shortest W1(5) nanowire. In addition, an upward shift of the bands
with increasing nanowire length is evident. Although the shift tends
to decrease gradually as the nanowires get longer, it could
potentially lead to a large difference in the transport properties
between the nanowires of different lengths.

\subsection{Spin disorder effect on the transport properties\label{ssec:res-sd}}

In the following, we use the terms ``spin ordered'' and ``collinear''
as synonyms. We make this clarification because the true ground state
at the interface of the TL systems is non-collinear but still spin
ordered in a checkerboard pattern (as we discussed in the beginning of
Sec.~\ref{ssec:res-est0}).

\begin{figure*}
  \includegraphics[width=\textwidth]{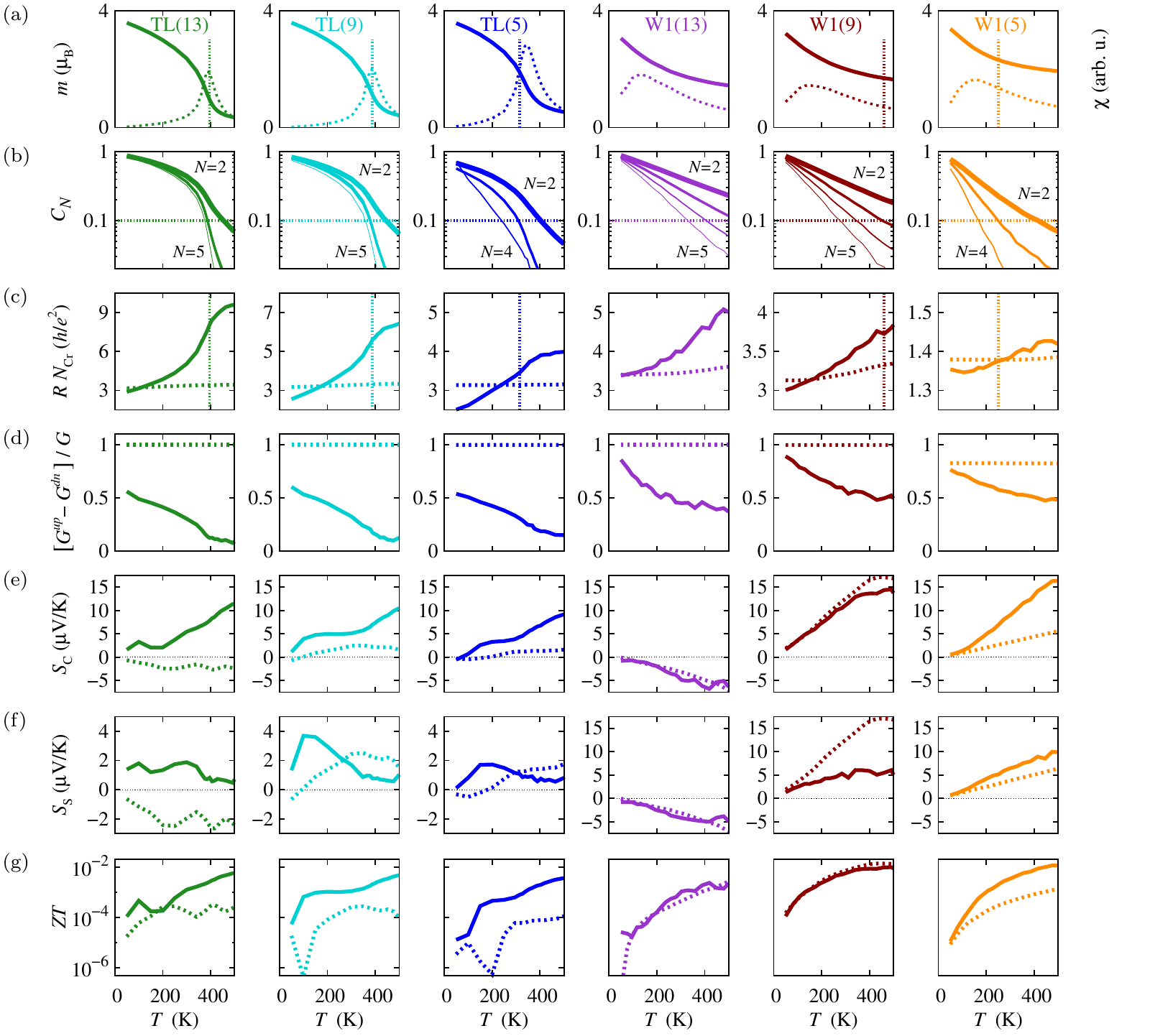}
  \caption{(Color online) (a) Monte Carlo site averages of the Cr
    atoms magnetic moment $m$ (solid line) and susceptibility $\chi$
    (dashed line). (b) Spatial correlation $C_N$ of the magnetic
    moment orientation between $N$-th nearest neighbor layers in the
    $z$ direction. Thick-to-thin line corresponds to second to $N$-th
    nearest neighbors, respectively. Vertical dotted line in (a) [and
    (c)] indicates temperature for which $C_3$ drops under 0.1
    [intersection with dotted line in (b)]. (c) Electrical resistance
    multiplied by the number of Cr atoms in the supercell
    cross-section (9 and 1 for TL and W1 systems, respectively). (d)
    Polarization of the electrical conductance. (e) Charge- and (f)
    spin-Seebeck coefficient. (g)~Thermoelectric figure of merit. In
    (c)-(g), dashed and solid line corresponds to the spin-ordered and
    spin-disordered data, respectively.}
  \label{fig:results}
\end{figure*}

Figure~\ref{fig:results}(a) shows the temperature dependence of the
average magnetization and the magnetic susceptibility $\chi(T)$ per
system. While the divergence of $\chi$ is associated with the critical
temperature in bulk systems, a falloff of $C_N$ with temperature
[Fig.~\ref{fig:results}(b)] is a good indicator of the loss of
magnetic order in thin layers and
nanowires.\cite{2014-04//kovacik/mavropoulos//blugel} Here it is
convenient to define a reasonably low threshold below which the $C_N$
associated with the long range magnetic order is considered small and
the correlation in a distance of $N$ Cr atoms is practically lost. The
choice of the threshold value is necessarily somehow arbitrary; we set
it to 0.1 for $C_{N\geq{3}}$, shown by a dotted horizontal line in
Fig.~\ref{fig:results}(b), for which the corresponding temperature of
the most bulk-like TL(13) system coincides with the susceptibility
peak.

The loss of long range order can be also identified as a kink in the
resistance--\emph{vs}.--temperature plot
[Fig.~\ref{fig:results}(c)]. Looking first at the TL(13) system, the
susceptibility peaks at ${T \approx 400~\text{K}}$. Although the
growth rate of $R(T)$ is reduced at this temperature, the resistance
keeps increasing. The not very pronounced kink at ${400~\text{K}}$ can
be associated with the decrease of $C_3$ under 0.1. Further increase
of $R(T)$ is consistent with a slow falloff of $C_2(T)$. Essentially
identical behavior can be seen in the case of the TL(9) system. For
the TL(5) system, the kink position is slightly shifted to a lower
temperature and the character of $C_4$, i.e., the correlation between
Cr moments in the two interface layers, is significantly changed. In
the case of nanowires, the susceptibility peak can not be associated
with any significant feature driven by the spin disorder. The average
magnetization stays sizeable up to high temperatures and no clear sign
of a kink in $R(T)$ is present. Although the ${C_3<0.1}$ condition is
fulfilled for the W1(5) and W1(9) systems in the examined temperature
range, the preservation of a strong short range order (see $C_2$)
masks any possible kink in $R(T)$.

Assuming that the CrTe layer became thicker and more bulk-like, it
would be interesting to evaluate the resistivity instead of the
resistance (practically excluding interface-resistance effects).  The
spin disorder contribution to the resistivity can be derived from the
slope of the linearly increasing resistance as a function of thickness
of the core region (lead-to-lead distance). We derive it from the
resistance values of the systems with 13 and 9 Cr layers at
${T=500~\text{K}}$ (where the spin disorder is almost saturated) as
${\rho_\text{sd}=N_\text{Cr}a_\text{lat}(R^{(13)}_\text{500~K}-R^{(9)}_\text{500~K})/4}$. Here,
$N_\text{Cr}$ is the number of Cr atoms in the supercell cross-section
(9 for TL and 1 for W1 systems). The calculation results in values of
12 and ${5~\upmu\Omega~\text{m}}$ for the CrTe slab and monoatomic
wire, respectively. The former value is in reasonable agreement with
the experimental estimate of the joint spin disorder and phonon
contribution to the resistivity (${20~\upmu\Omega~\text{m}}$) of bulk
Cr$_{0.9}$Te in the NiAs structure measured at
${400~\text{K}}$,\cite{1989-11//dijkstra/weitering//de-groot}$^,$\footnote{The
resistivity measurements in
Ref.~\onlinecite{1989-11//dijkstra/weitering//de-groot} were reported
up to ${400~\text{K}}$, which was already above the Curie
temperature.} as well as with the more recent measurement of the CrTe
thin film resistivity at RT
(${\approx{10}~\upmu\Omega~\text{m}}$).\cite{2007-09//sreenivasan/teo//osipowicz}

\begin{figure}[!h]
  \includegraphics[width=\columnwidth]{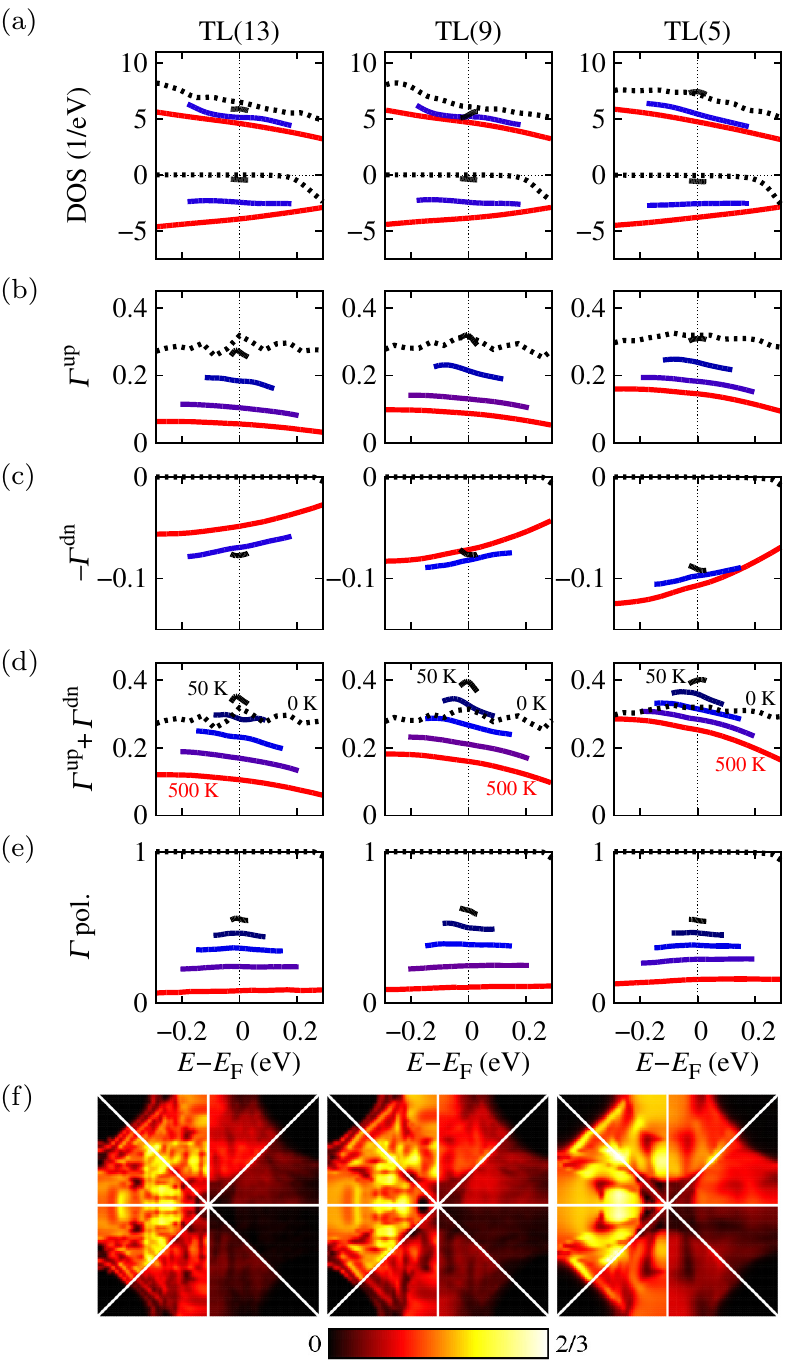}
  \caption{(Color online) (a) DOS at the Cr atoms in the central layer
  of the TL systems (positive and negative values correspond to the
  $\uparrow$ and $\downarrow$ spin, respectively). The transmission
  probability $\varGamma$ for (b) $\uparrow$ and (c) $\downarrow$
  spin, (d) total transmission probability and (e) its polarization
  ${(\varGamma^\uparrow-\varGamma^\downarrow)/(\varGamma^\uparrow+\varGamma^\downarrow)}$. Dotted
  and solid line in (a)-(e) correspond to the spin ordered and
  disordered data, respectively, the energy range of spin disordered
  data corresponds to ${\pm 7k_\text{B}T}$ around $E_\text{F}$, where
  $T$ is the MC simulation temperature. (f) The transmission
  probability as a function of $\mathbf{k}_\parallel$ within the
  surface Brillouin zone in the systems with ${1 \times 1}$ unit cell
  cross-section. The irreducible wedges of the SBZ in the upper and
  lower half of the plots represent
  ${\varGamma^\uparrow+\varGamma^\downarrow}$ and
  ${\varGamma^\uparrow-\varGamma^\downarrow}$, respectively, while
  from left to right the wedges correspond to the collinear case and
  the temperatures ${100~\text{K}}$, ${300~\text{K}}$ and
  ${500~\text{K}}$.}
  \label{fig:dostrantl}
\end{figure}

\begin{figure}
  \includegraphics[width=\columnwidth]{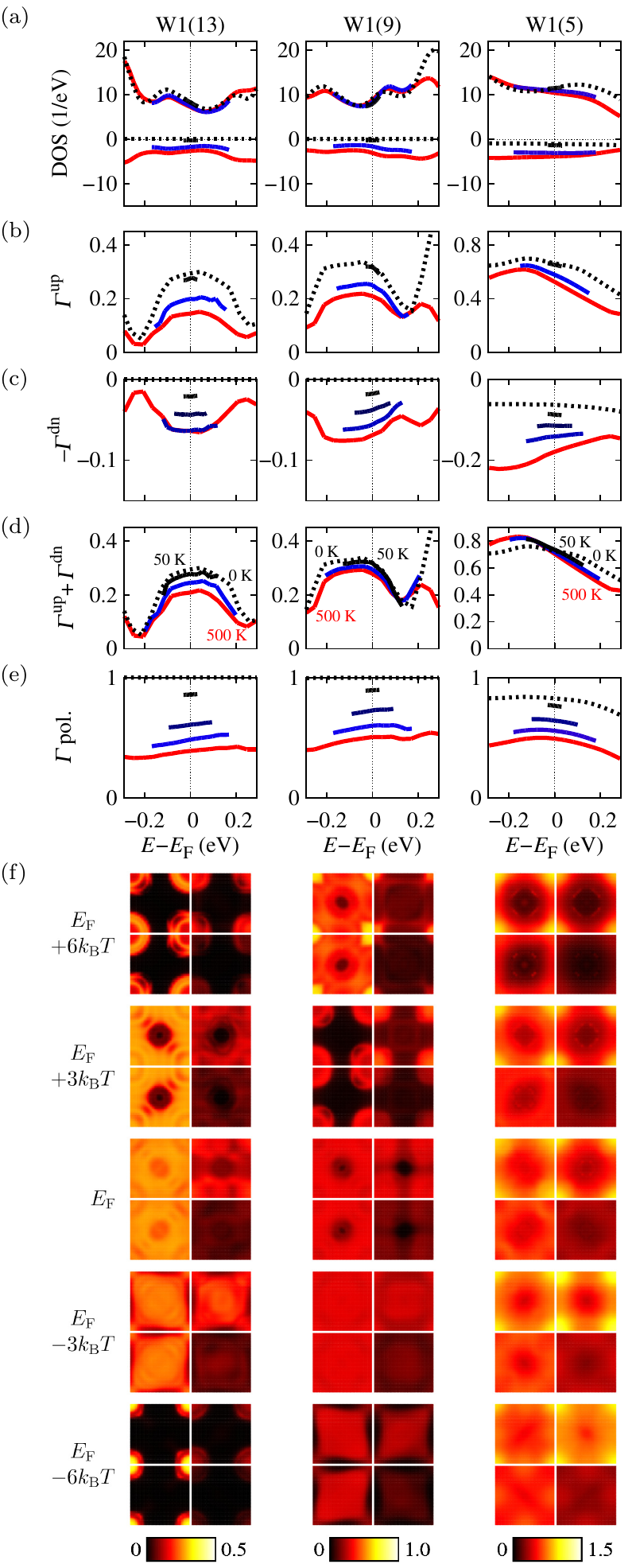}
  \caption{(Color online) DOS at the central layer Cr atom of the W1
  systems. For (a)-(e) plots, the corresponding description in
  Fig.~\ref{fig:dostrantl} is valid. (f) The
  ${\varGamma^\uparrow+\varGamma^\downarrow}$ and
  ${\varGamma^\uparrow-\varGamma^\downarrow}$ are shown in the upper
  and lower half of the plots, respectively, as a function of the
  energy around $E_\text{F}$ and $\mathbf{k}_\parallel$ within the
  surface Brillouin zone. The left and right part of the SBZ
  correspond to the temperature of ${0~\text{K}}$ and
  ${500~\text{K}}$, respectively.}
  \label{fig:dostranw1}
\end{figure}

In order to further analyze the transport properties, we present the
energy dependent and spin resolved DOS in the core region central
layer and the transmission probability $\varGamma$, as well as the
$\mathbf{k}_\parallel$-resolved $\varGamma$, for the thin layers in
Fig.~\ref{fig:dostrantl} and for the nanowires in
Fig.~\ref{fig:dostranw1}. The up and down spin projected transmission
probabilities are calculated from the individual matrix elements as
$\varGamma^\uparrow=\varGamma^{\uparrow\uparrow}+(\varGamma^{\uparrow\downarrow}+\varGamma^{\downarrow\uparrow})/2$
and
$\varGamma^\downarrow=\varGamma^{\downarrow\downarrow}+(\varGamma^{\uparrow\downarrow}+\varGamma^{\downarrow\uparrow})/2$,
respectively. All data sets in the graphs as a function of energy are
plotted as solid lines in the energy range of $E_\text{F}\pm
7k_\text{B}T$ in the case of spin disorder, while dotted lines in the
energy range ${E_\text{F}\pm 7k_\text{B}\times 500~\text{K}}$ are used
for the spin ordered (magnetic collinear) case. Although it is well
established that DOS features are not sufficient to determine the
character of the transmission probability, we examine first whether
such connection could be found in our systems. For brevity, we refer
to the majority and minority spin channel as $\uparrow$ and
$\downarrow$ (see Sec.~\ref{ssec:met-sdet} for the spin axis
convention).

The DOS of all TL systems is very similar, not only at the
${T=0~\text{K}}$ calculations but also at elevated temperatures
[Fig.~\ref{fig:dostrantl}(a)]. The DOS$^\uparrow$ at ${0~\text{K}}$
exhibits an asymmetry around $E_\text{F}$ which is not reflected in
the character of $\varGamma^\uparrow(E)$ at 0~K
[Fig.~\ref{fig:dostrantl}(b)]. The bottom edge of the Cr $d$ minority
spin band that can be seen in the DOS$^\downarrow$ at 0~K for
${E-E_\text{F}>{0.2}~\text{eV}}$ manifests itself only very weakly in
$\varGamma^\downarrow(E)$ [Fig.~\ref{fig:dostrantl}(c)]. The spin
disorder leads to almost equal DOS$^\uparrow$ and DOS$^\downarrow$
above $T_\text{c}$ as expected, but the onset of the spin mixing at
low temperatures is very slow. The effect of spin disorder on
$\varGamma^\uparrow$ has also a slow onset at low $T$ but gradually a
clear asymmetry around $E_\text{F}$ develops, consistently with the
DOS$^\uparrow$. The $\varGamma^\uparrow$ suppression due to the spin
disorder is obviously proportional to the thickness of the core region
(i.e., the system is in the Ohmic regime at high $T$).

Very different picture can be seen for $\varGamma^\downarrow$, which
displays an immediate fast onset at low $T$ that then remains almost
temperature independent. This effect originates from the Cr atoms at
the interface layer favoring the non-collinear ground-state
configuration described in Sec.~\ref{ssec:res-est0}. Since the
self-consistent electronic structure was calculated in the collinear
state and the spin-ordered calculations refer to this state, they show
a suppression of $\varGamma^\downarrow$ by the half-metallic gap.

When accounting for the non-collinear state at the interface at ${T
\rightarrow 0}$ (where the interior of the spacer is half-metallic),
spin-flip processes occur, during which an incoming spin-down electron
from the lead is flipped at the first interface to spin-up, passes the
magnetic spacer and may or may not be flipped back to spin down at the
second interface (where again the moments form a non-collinear
state). Thus $\varGamma^\downarrow$ can be non-zero even if the
central part of the spacer is perfectly half-metallic.

Overall, this effect leads to an increased conductance (decreased
resistance) at low $T$ as can be seen in Figs.~\ref{fig:dostrantl}(d)
and \ref{fig:results}(c), especially pronounced in the thin quasi-2D
TL(5) system.  While it may, in general, go unnoticed when looking at
the resistance of thicker systems, the effect leaves an almost
identical fingerprint in the conductance polarization at low $T$ for
all different TL systems irrespective of their thickness
[Figs.~\ref{fig:results}(d) and \ref{fig:dostrantl}(e)]. Further
temperature increase affects the conductance polarization
[Fig.~\ref{fig:results}(d)] of the TL systems also in a similar way. A
weakly pronounced kink at $T_\text{c}$ can be seen, above which the
polarization is negligible. The transmission probability as a function
of $\mathbf{k}_\parallel$ is shown in Fig.~\ref{fig:dostrantl}(f) only
for $E_\text{F}$, since no qualitative differences can be observed in
the considered energy range. In this figure, each triangular region
corresponds to the irreducible part of the SBZ for different
temperatures and for the sum or difference of $\varGamma^\uparrow$ and
$\varGamma^\downarrow$, as described in the caption. The region with
no conductance around the M point slightly enlarges with increasing
$T$. The fine structure of both $\varGamma$ and its polarization can
be still recognized at ${T=100~\text{K}}$, while it is almost
completely smeared out due to the spin disorder at the room
temperature.

The energy dependence of the DOS and $\varGamma$ is much richer in
case of the monoatomic wires. The spin disorder has, in general, a
small effect on the DOS [Fig.~\ref{fig:dostranw1}(a)] and even at
${T=500~\text{K}}$, the DOS$^\downarrow$ is significantly lower than
the DOS$^\uparrow$. The gradual decrease of $\varGamma^\uparrow$ as a
consequence of the spin disorder [Fig.~\ref{fig:dostranw1}(b)] is
accompanied by the corresponding increase of $\varGamma^\downarrow$
[Fig.~\ref{fig:dostranw1}(c)]. The resulting total $\varGamma$ is
therefore only weakly temperature dependent
[Fig.~\ref{fig:dostranw1}(d)]. The character of $\varGamma$ as a
function of energy seems to be completely uncorrelated with the DOS,
with a possible exception in the peak at
${E-E_\text{F}={0.25}~\text{eV}}$ for the W1(9) system, which is
strongly suppressed in both $\varGamma$ and DOS. Only the W1(5) system
loses half-metallicity at ${T=0~\text{K}}$, possibly due to a
penetration of Ag states throughout the core region.

The conductance polarization [Figs.~\ref{fig:results}(d) and
\ref{fig:dostranw1}(e)] decreases gradually with temperature similarly
to the total magnetization [Fig.~\ref{fig:results}(a)], remaining
sizable at ${T=500~\text{K}}$. It is worth to note that an energy
shift of about ${0.15~\text{eV}}$ between the characteristic
$\varGamma$ peak of the W1(13) and W1(9) systems is very similar to
the energy shift observed in the corresponding structure of
${\mathbf{k}_\parallel\text{-}}$resolved DOS
[Fig.~\ref{fig:qdos}(c)]. This shift of about $3k_\text{B}T$ can be
also identified in the $\mathbf{k}_\parallel$-resolved
$\varGamma$. While the W1(13) and W1(9) systems exhibit rather rich
fine structure of $\varGamma(\mathbf{k}_\parallel)$, the features of
$\varGamma(\mathbf{k}_\parallel)$ are mostly smeared out in the case
of the W1(5) system [Fig.~\ref{fig:dostranw1}(f)].

Now we proceed to discuss the charge- and spin-Seebeck coefficient
shown in Figs.~\ref{fig:results}(e) and \ref{fig:results}(f),
respectively. The charge-Seebeck coefficient $S_\text{C}$ of the TL
systems in the collinear magnetic case is very low owing to no
significant asymmetry in $\varGamma$ [dotted line in
Fig.~\ref{fig:dostrantl}(d)]. The spin disorder has a very similar
effect for all TL systems, $S_\text{C}$ is positive [negative slope of
$\varGamma$ in Fig.~\ref{fig:dostrantl}(d)] and grows with increasing
temperature. The corresponding spin-Seebeck coefficient $S_\text{S}$
also becomes consistently positive due to the spin disorder but is
almost completely suppressed already at the room temperature. No
systematic trend on the nanowire length can be observed for the
Seebeck coefficients given the strongly changing character of
$\varGamma$ among the nanowires. Also, while the spin disorder
increases both $S_\text{C}$ and $S_\text{S}$ in the case of W1(5), it
induces their decrease for the W1(9) or has just a small effect in the
case of W1(13). In Fig.~\ref{fig:results}(g), we show the resulting
figure of merit ($ZT$). Clearly, its value in all cases is still too
low for practical applications, however, the role of spin disorder in
its enhancement [except the W1(9) and W1(13) systems] is noteworthy.

\subsection{Substitutional impurity effect on the transport properties of W1(13) nanowire}\label{ssec:res-imp}

\begin{figure*}
  \includegraphics[width=\textwidth]{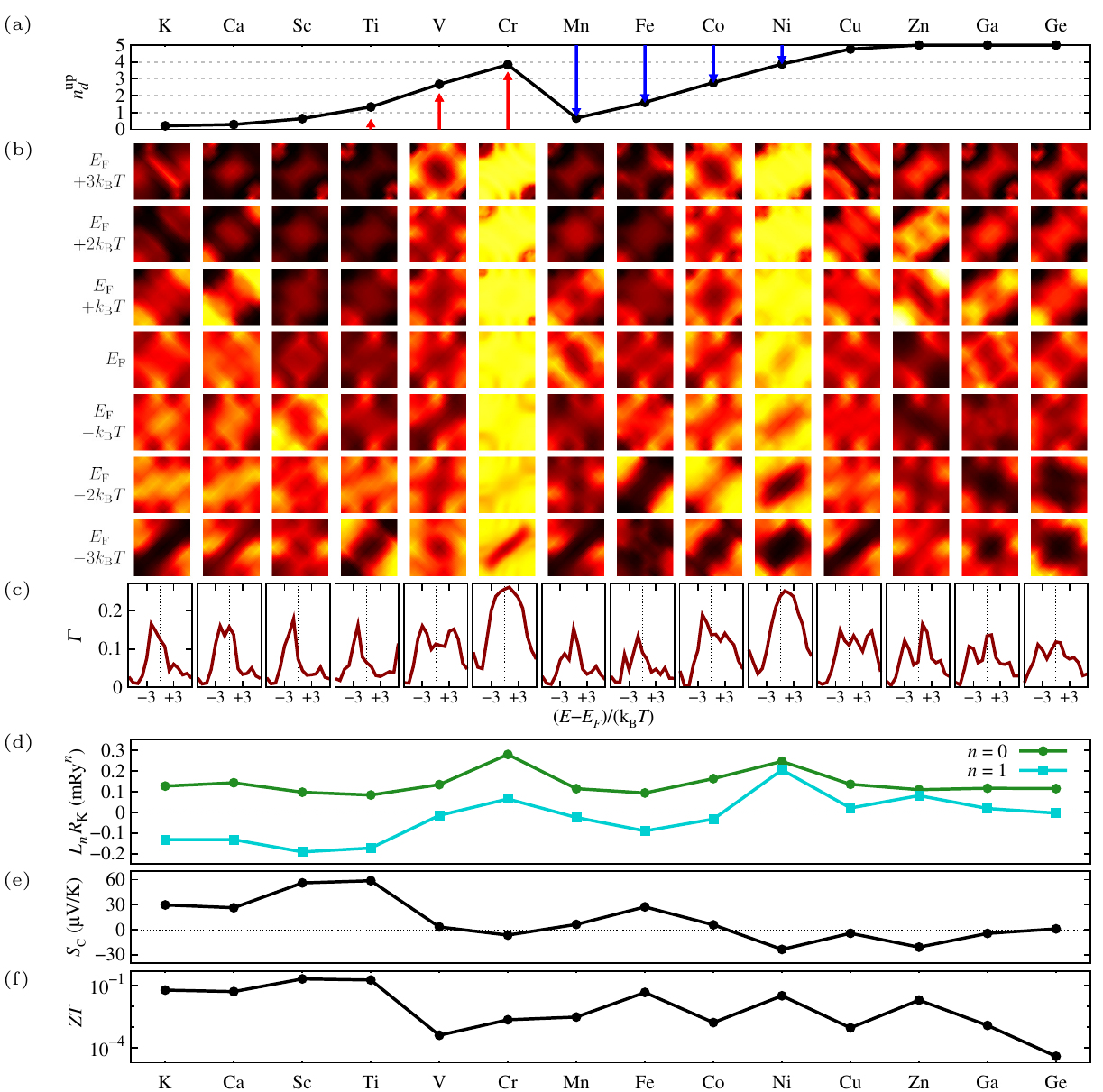}
  \caption{(Color online) Properties of the W1(13) nanowire with the
  substitutional impurity in its central layer. (a) Majority spin $d$
  orbitals occupation of the impurity site (solid line). Magnetic
  moment m($\mu_\text{B}$) of the impurity site (if
  ${m>0.4~\mu_\text{B}}$) is represented by a vertical arrow, pointing
  up/down in case of parallel/antiparallel orientation with the
  nanowire magnetization. (b) The transmission probability as a
  function of the energy around $E_\text{F}$ and
  $\mathbf{k}_\parallel$ within a section of the surface Brillouin
  zone (double the irreducible part) defined by $\Gamma$ (bottom left
  and top right corner) and M (bottom right and top left corner) high
  symmetry points, with X point in the center. The gradient color code
  [equivalent to Figs.~\ref{fig:dostrantl}(f) and
  \ref{fig:dostranw1}(f)] corresponds to a linear range of $\varGamma$
  between 0 and 0.37. (c) The $\mathbf{k}_\parallel$-integrated
  transmission probability as a function of the energy around
  $E_\text{F}$. (d) The transport coefficients $L_0$ and $L_1$
  (${R_\text{K}=h/e^2}$ is the von Klitzing constant). (e) Charge
  Seebeck coefficient. (f)~Thermoelectric figure of merit.}
  \label{fig:imp}
\end{figure*}

The strong modulation of $\varGamma(E)$ in the case of the nanowires
motivated us to further investigate the effect of a substitutional
impurity, as a source of extra scattering, on the transport properties
of the nanowire. We chose the W1(13) nanowire as a representative
system. The main criterion was the robustness of the Seebeck
coefficients of W1(13) with respect to the spin disorder. Furthermore,
an energy shift or a shape modulation of the $\varGamma(E)$ peak
around $E_\text{F}$ could lead to a large enhancement of the Seebeck
coefficients as a result of an arising asymmetry. In the W1(13)
nanowire, the Cr atom in the wire center was substituted by an element
of the fourth period, systematically from potassium to germanium. The
electronic structure of each nanowire with the substitutional impurity
was calculated self-consistently using the impurity Green function
method.\cite{bauer-thesis,2002-03//papanikolaou/zeller/dederichs} Due
to the already mentioned Seebeck coefficient robustness with respect
to the spin disorder, we first calculated the transport properties for
the spin ordered case with a temperature of ${470~\text{K}}$ entering
via the Fermi smearing, the main results of which are summarized in
Fig.~\ref{fig:imp}.

A sizable magnetic moment at the impurity site was stabilized for all
elements from Ti to Ni [Fig.~\ref{fig:imp}(a)]. While the magnetic
moments of Ti and V impurities were aligned in parallel with the
overall nanowire magnetization, the anti-parallel alignment was
energetically much more favorable for the Mn, Fe, Co and Ni
impurities. The resulting majority spin $d$ character occupation of
the Mn--Ni sequence is therefore qualitatively similar to the Sc--Cr
one. The $\mathbf{k}_\parallel$-resolved transmission probability is
shown in Fig.~\ref{fig:imp}(b) at several energy values around
$E_\text{F}$. The character of $\varGamma(\mathbf{k}_\parallel)$ does
not only look remarkably alike for the wires with the impurity $d$
shell being essentially empty (K and Ca) or fully occupied (Ga and Ge)
but also for the pairs of the transition metal impurities with a
similar majority spin $d$ character occupation (with a possible energy
shift of the prominent features); namely, Sc is alike to Mn, Ti to Fe,
V to Co, Cr (no impurity) to Ni. The
${\mathbf{k}_\parallel\text{-}}$integrated transmission probability
$\varGamma(E)$ around $E_\text{F}$ depicted in Fig.~\ref{fig:imp}(c)
indeed shows that certain impurities lead to a significant asymmetry
of $\varGamma$ at $E_\text{F}$. Apart from the Ni impurity, which
causes a shape preserving energy shift of $\varGamma$ [compared to the
original W1(13) wire], all other impurities lead to a strong
$\varGamma$ modulation. The transport coefficient $L_1$ as a measure
of the $\varGamma$ asymmetry around $E_\text{F}$ is strongly enhanced
in case of Sc, Ti and Ni impurities while the $L_0$ coefficient is
generally reduced by all impurities except Ni
[Fig.~\ref{fig:imp}(d)]. The Sc and Ti impurities thus lead to an
overall enhancement of $S_\text{C}$ by an order of magnitude in
comparison with the W1(13) nanowire [Fig.~\ref{fig:imp}(e)] and two
orders of magnitude increase of $ZT$, reaching values of about 0.2.

The spin disorder influence on the transport properties was examined
in the case of Sc and Ti impurities at room (${290~\text{K}}$) and
elevated (${470~\text{K}}$) temperatures, with results summarized in
Table~\ref{tab:imp}. While the $\varGamma(E)$ shape is affected only
weakly by the spin disorder, the slightly suppressed conductance leads
to a further enhancement of the charge Seebeck coefficient. For the Sc
impurity, an additional increase at room temperature leads to an
$S_\text{C}$ of ${80~\upmu\text{V}/\text{K}}$ and a corresponding $ZT$
of 0.35. The spin Seebeck coefficient is reduced to about 45\% of the
respective $S_\text{C}$ value due to the spin disorder, remaining
sizable even at elevated temperatures.

\begin{table}
  \caption{Charge and spin Seebeck coefficients (in
  ${\upmu\text{V}/\text{K}}$) and figure of merit for the W1(13)
  nanowire with selected substitutional impurities (Sc and Ti). Note
  that in the spin ordered case, ${S_\text{C} \approx S_\text{S}}$ due
  to negligible $L_n^{\downarrow\downarrow}$,
  $L_n^{\uparrow\downarrow}$ and $L_n^{\downarrow\uparrow}$ terms in
  comparison to $L_n^{\uparrow\uparrow}$ for both ${n=0}$ and
  ${n=1}$.}
  \begin{ruledtabular}
    \begin{tabular*}{\columnwidth}{@{\extracolsep{\fill}}lccccc}
      & \multicolumn{2}{c}{spin order} &
      \multicolumn{3}{c}{spin disorder}\\\cline{2-3}\cline{4-6}
      impurity &
      ${S_\text{C} \approx S_\text{S}}$ & $ZT$ &
      $S_\text{C}$ & $S_\text{S}$ & $ZT$\\\hline
      Sc (290 K) & 71 & 0.27 & 80 & 38 & 0.35 \\
      Sc (470 K) & 56 & 0.21 & 58 & 25 & 0.19 \\
      Ti (290 K) & 47 & 0.09 & 58 & 26 & 0.12 \\
      Ti (470 K) & 59 & 0.18 & 64 & 28 & 0.18 
    \end{tabular*}
  \end{ruledtabular}
  \label{tab:imp}
\end{table}

\section{Summary and conclusions\label{sec:sum}}

We have modeled the electron and spin transport properties through
CrTe nanostructures (thin layers and monoatomic wires) at elevated
temperatures. We focused on the effects of spin disorder, i.e., on the
effects of the fluctuating magnetic moments at ${T>0}$.

Our calculations show the importance of these effects, both
quantitatively and qualitatively. Examining the resistance, current
polarization, charge-Seebeck and spin-Seebeck coefficients in a number
of structures and temperatures, we find that they depend on system
dimensionality (thin layer vs. nanowire), system size (film thickness
or wire length) and temperature. Simplifying the calculation by
accounting for the electronic temperature (i.e., the Fermi smearing)
alone is in many cases not even qualitatively adequate for a
description of the transport coefficients; but there are notable
exceptions, e.g. in the Seebeck coefficients and $ZT$ of the longer
(13-atom) nanowire. Unfortunately, nothing indicates the validity of
this simplified and computationally less expensive approach
beforehand; it can be verified only \emph{a posteriori}, after the
full calculation with spin disorder.

A few general observations can be made about our results. In the thin
films, the crossover temperature $T_\text{c}$ (at the magnetic
susceptibility peak, signaling the strongest magnetic fluctuations)
coincides with a change of slope in the
resistance-\emph{vs.}-temperature curve, as is known from bulk
systems. However, this is not the case in the nanowires, where also
the susceptibility peak is less pronounced and the magnetization drop
is much smoother. The difference stems possibly from the fact that the
fluctuations at $T_\text{c}$ in the wires are long-ranged (due to
their lower dimensionality) compared to the films, causing a smoother
gradient of magnetization (e.g., infinitely long wires show long-range
critical fluctuations at ${T_\text{c}=0}$). As $T$ grows beyond
$T_\text{c}$, the fluctuations in wires become gradually short-ranged,
increasing the resistance. In films, $T_\text{c}$ is at a point where
the fluctuations are already short-ranged (as can be seen from the
correlation functions), so that an increase of $T$ does not produce
proportionally more scattering; thus the kink is created.

A second observation is that, although the current spin polarization
seems to consistently drop with increasing spin disorder, this is not
always the case for the spin Seebeck coefficient, especially for the
nanowires. At the same time, the current spin polarization of the
nanowires is not fully suppressed even at high temperatures. The
reason again lies in the long-ranged correlations in the
nanowires. The magnetization of a short nanowire behaves to an extent
as a macro-spin, i.e., a superparamagnetic entity, so that the
fluctuations play a smaller role. Obviously, at very high temperatures
or for very long nanowires the transmission as a function of energy
must become spin-independent, suppressing the spin-Seebeck
coefficient. However, as presented in this study, this is not the case
at moderate temperatures or short lengths.

A third observation is that $ZT$ may increase by orders of magnitude
either by the effect of spin disorder or if an impurity is placed in
the middle of the nanowire. The effect of a substitutional impurity
strongly depends on its type, with Sc and Ti being the best candidates
through the $3d$ series for an increased $ZT$.

Finally, our results contribute to the question on the usefulness of
half-metallic ferromagnets in order to achieve spin polarized electron
transport. We see from the thin-film calculations that if the
interface of a half-metallic magnet to the leads develops a
non-collinear magnetic phase, then the current spin polarization drops
drastically: under these conditions, the half metallic character in
the interior does not improve the current spin polarization compared
to the values of a regular ferromagnet. Additionally, non-collinear
states at ${T>0}$ due to local-moment fluctuations, further reduce the
current spin polarization. We also verify the (expected) result that
even in the absence of fluctuations, half-metallicity does not imply a
strong spin-Seebeck effect, since the same spin channel can contribute
positively or negatively to the spin-Seebeck coefficient.

\acknowledgments

We are indebted to Daniel Wortmann, Voicu Popescu and Carmen
E. Quiroga for enlightening discussions. Support from the Deutsche
Forschungsgemeinschaft (SPP 1538 ``Spin Caloric Transport'') is
gratefully acknowledged. Computational resources were provided by the
JARA-HPC from the RWTH Aachen University under project jara0051.

\bibliography{references}

\end{document}